\documentclass{iopjournal}
\usepackage{graphicx,empheq,amssymb,orcidlink,bm}
\usepackage{amsmath}
\usepackage{graphicx}
\usepackage{dcolumn}
\usepackage{mathrsfs}
\usepackage{amssymb}
\usepackage{amsfonts,multirow}
\usepackage{bm,lineno}
\usepackage{color}

\begin{document}

\articletype{Paper} 

\title{Converting $PT$-Symmetric Topological Classes by Floquet Engineering}

\author{Ming-Jian Gao$^{1,*}$\orcid{0000-0002-6128-8381} and Jun-Hong An$^{1}$\orcid{0000-0002-3475-0729}}

\affil{$^1$Key Laboratory of Quantum Theory and Applications of Ministry of Education, Lanzhou Center for Theoretical Physics, Gansu Provincial Research Center for Basic Disciplines of Quantum Physics, Key Laboratory of Theoretical Physics of Gansu Province, Lanzhou University, Lanzhou 730000, China}

\affil{$^*$Author to whom any correspondence should be addressed.}

\email{120220908581@lzu.edu.cn}

\keywords{Topological insulator, Floquet engineering, $PT$ symmetry, Topological phase transition}

\begin{abstract}
Going beyond the conventional classification rule of Altland-Zirnbauer symmetry classes, $PT$ symmetric topological phases are classified by $(PT)^2=1$ or $-1$. The interconversion between the two $PT$-symmetric topological classes is generally difficult due to the constraint of $(PT)^2$. Here, we propose a scheme to control and interconvert the $PT$-symmetric topological classes by Floquet engineering. We find that it is the removal of the $\mathbb{Z}_2$ gauge degree of freedom, induced by the $\pi$ phase difference between different hopping rates, by the periodic driving that leads to such an interconversion. Relaxing the system from the constraint of $(PT)^2$, rich exotic topological phases, e.g., the coexisting $PT$-symmetric first-order real Chern insulator and second-order topological insulator not only in different quasienergy gaps, but also in one single gap, are generated. In contrast to conventional Floquet topological phases, our result provides a way to realize exotic topological phases without changing symmetries. It enriches the family of topological phases and gives an insightful guidance for the development of multifunctional quantum devices.
\end{abstract}

\section{Introduction}
As an indispensable and significant field in modern physics, topological phases that go beyond Landau symmetry-breaking theory not only enrich the paradigm of condensed matter physics, but also provide new directions for the development of quantum technology \cite{RevModPhys.82.3045, RevModPhys.83.1057,RevModPhys.87.137,RevModPhys.88.035005, RevModPhys.90.015001, RevModPhys.93.025002}. Including topological insulator \cite{RevModPhys.82.3045,RevModPhys.83.1057}, superconductor \cite{RevModPhys.83.1057,RevModPhys.87.137}, and semimetal \cite{RevModPhys.90.015001,RevModPhys.93.025002}, topological phases are signified by the formation of symmetry-protected boundary states and can be characterized by topological invariants of the bulk energy bands. This, as one of the most significant features of topological phases, is called bulk-boundary correspondence. Symmetries play a dominant role in the classification of topological phases. Through three internal time-reversal, chiral, and particle-hole symmetries, we can classify topological phases into tenfold Altland-Zirnbauer symmetry classes \cite{RevModPhys.88.035005,RevModPhys.88.021004}. This has become the cornerstone for the development of topological phases.

It was recently found that the three internal symmetries are insufficient to classify all topological phases. Crystal symmetry also plays a crucial role in many topological systems \cite{doi:10.1126/science.aah6442,PhysRevLett.129.046802,PhysRevLett.128.026405,PhysRevX.7.041069}. In particular, the topological phases with the combined space-time inversion ($PT$) symmetry do not obey the Altland-Zirnbauer classification rule \cite{RevModPhys.91.015006,RN237,PhysRevLett.125.053601,Guo_2021,Jiang2021,Peng2022,Bouhon_2020,Slager_2024}. The inversion symmetry in the presence of the $\mathbb{Z}_2$ gauge degree of freedom induced by the hopping amplitudes with a $\pi$ phase difference satisfying the $\mathbb{Z}_2$ gauge conditions should be projectively represented, which alters the algebraic structure of the symmetries and thus generates rich exotic topological phases \cite{PhysRevB.102.161117}. The projectively represented inversion operator makes the symmetry classes change from $(PT)^2=1$ to $-1$ in spinless system and from $(PT)^2=-1$ to $1$ in spin-1/2 system. The $PT$-symmetric topological phases can be roughly divided into two categories with $(PT)^2=\pm 1$ \cite{PhysRevLett.116.156402}. 
\begin{figure}[tbp]
\centering
\includegraphics[width=\columnwidth]{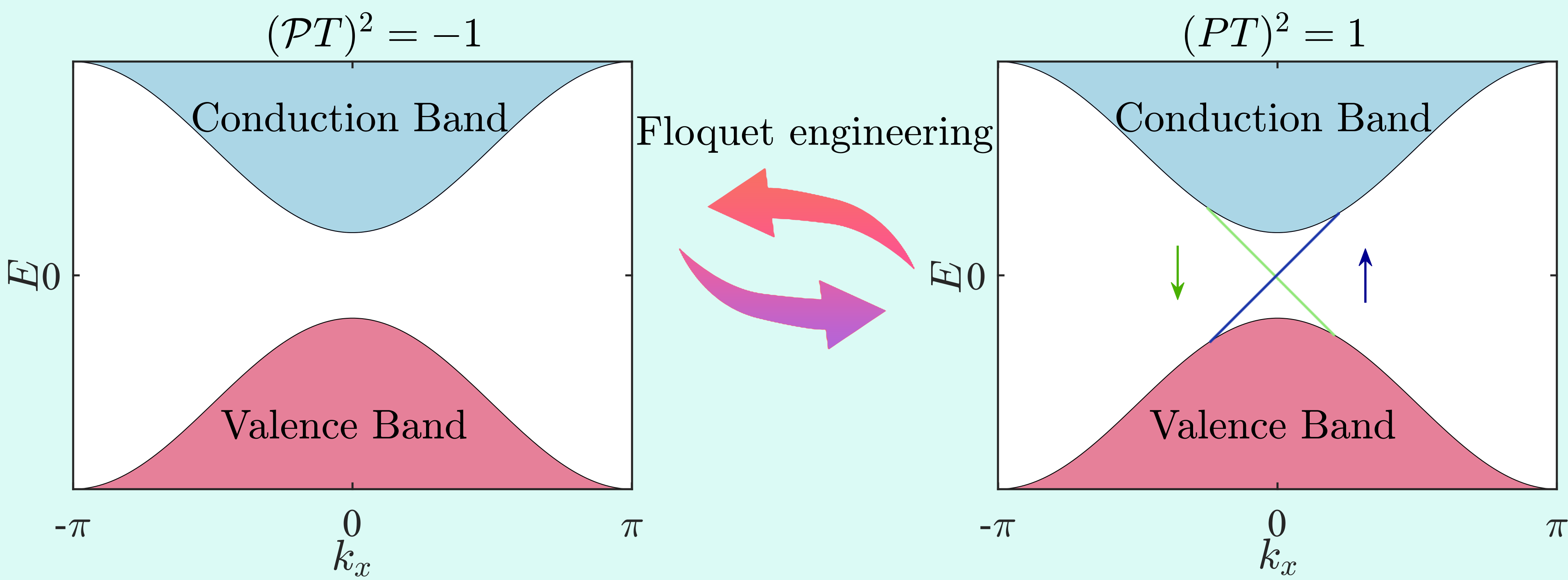}
\caption{Converting $PT$-symmetric topological classes by Floquet engineering. Two-dimensional $\mathcal{P}T$-symmetric systems with $(\mathcal{P}T)^2=-1$ does not have the gapless edge states. The one with $(PT)^2=1$ allows the gapless edge states.}\label{fig:1}
\end{figure}
The systems with $(PT)^2=-1$ host the M\"obius topological insulator, in which the boundary spectra are twisted in momentum space \cite{PhysRevLett.128.116803,PhysRevLett.128.116802,PhysRevB.102.161117}, and the axion insulator, which exhibits a quantized magnetoelectric response \cite{Peizhe_2016,Qiu_2023}. Possessing real Bloch wave functions, the $PT$-symmetric systems with $(PT)^2=1$ host the real Chern insulator \cite{PhysRevLett.125.126403,PhysRevLett.125.053601,Guo_2021,Jiang2021,Peng2022,Bouhon_2020,Slager_2024}, which is characterized by a $\mathbb{Z}_2$ topological invariant of the Stiefel-Whitney class \cite{PhysRevX.8.031069,PhysRevLett.121.106403}, and the Euler topological phases, which have multigap topologies with band nodes carrying non-Abelian charges \cite{PhysRevLett.125.053601,Guo_2021,Jiang2021,Peng2022,Bouhon_2020,Slager_2024}. These advances indicate that spatial inversion symmetry opens a door to the discovery of exotic topological phases. However, limited by the value of $(PT)^2$, the interconversion among these $PT$-symmetric topological phases is difficult. In practical application in designing multiplexing devices, it is generally desirable to exploit the respective advantages of the different classes of topological phases. A natural question is how to realize the interconversion between different $PT$-symmetric topological classes on demand. On the other hand, Floquet engineering has become a useful tool to create topological phases due to its distinguished role in changing the symmetries of systems \cite{PhysRevLett.121.036401,PhysRevLett.123.016806,PhysRevLett.124.057001,PhysRevLett.124.216601,PhysRevLett.117.087402,McIver2020,Fan2024}. Can Floquet engineering interconvert the topological classes, while still preserving the $PT$ symmetry?

Addressing these problems, we here propose a scheme to convert the $PT$-symmetric topological class from $(PT)^2=-1$ to $1$ by Floquet engineering, see Fig. \ref{fig:1}. Our analysis reveals that it is the removal of the $\mathbb{Z}_2$ gauge degree of freedom that makes such a conversion realizable. Setting free from the constraint of $(PT)^2$, exotic $PT$-symmetric two-dimensional topological phases with coexisting first- and second-order topological insulators are realized not only in different quasienergy gaps but also in one single gap. They are completely absent in static $PT$-symmetric systems. Going beyond the conventional Floquet topological phases, our result lays a foundation to realize exotic topological phases without changing symmetries. The coexisting topological phases have the potential to facilitate the design of multifunctional quantum devices.

\section{Floquet $PT$-symmetric topological phases}
The Altland-Zirnbauer symmetry classes governed by three internal symmetries do not exhaust the topological phases \cite{PhysRevLett.116.156402,PhysRevLett.125.126403,PhysRevLett.126.196402,PhysRevLett.128.116803,PhysRevLett.130.026101}. It was found that some topological phases are determined not only by the internal symmetries but also by the spatial symmetry. The topological phases with the combined symmetries of time reversal $T$ and spatial inversion $P$ are classified by whether $(PT)^2=1$ or $-1$ and the difference between the system dimension and the codimension of the defect \cite{PhysRevLett.116.156402,RevModPhys.88.035005,PhysRevLett.126.196402}. From this classification \cite{PhysRevLett.116.156402,PhysRevLett.126.196402}, $PT$-symmetric system can realize real Chern insulator only in $(PT)^2=1$ case \cite{PhysRevLett.125.126403,PhysRevLett.126.196402}, second-order topological insulator in $(PT)^2=1$ case \cite{PhysRevLett.123.256402} and $(PT)^2=-1$ case \cite{doi:10.1126/science.aah6442}. Generally, $(PT)^2=1$ in spinless systems and $-1$ in spin-1/2 systems \cite{PhysRevLett.116.156402,PhysRevLett.126.196402}. On the other hand, in the field of topological physics, if the hopping rates between different sublattices have a $\pi$ phase difference, see the dashed lines and the solid lines in the same color in Fig. \ref{fig:2}(a), and there is a gauge transformation operator $G$ satisfying 
\begin{equation}\label{GPT}
[G,T]=0,\qquad   \{G,P\}=0,\qquad   G^2=1,
\end{equation}
then an additional $\mathbb{Z}_2$ gauge degree of freedom is exerted in the system. The $\mathbb{Z}_2$ gauge degree of freedom has been found to play an important role in the classification of the $PT$-symmetric topological phases \cite{PhysRevLett.126.196402}. It makes the $PT$ operation be projectively represented. This projective $PT$ symmetry was first proposed in Ref. \cite{PhysRevLett.126.196402}. According to the $\mathbb{Z}_2$ gauge theory, actual inversion operation becomes $\mathcal{P}=GP$ and we can find
\begin{equation}
    (\mathcal{P}T)^2=GPTGPT=-G^2PTPT=-(PT)^2.
\end{equation}
Thus, the $\mathbb{Z}_2$ gauge degree of freedom facilitates the realization of topological phases with $({\mathcal{P}T})^2=-1$ in spinless systems and topological phases with $({\mathcal{P}T})^2=1$ in spin-$1/2$ systems, which are opposite to those in gauge-free systems. It is noted that the prerequisite to projectively represent the inversion operator is that Eq. \eqref{GPT} must hold. The presence of the $\mathbb{Z}_2$ gauge field does not necessarily destroy the original $P$ symmetry \cite{PhysRevLett.126.196402}. This means that there are several inversion symmetry operators under which the system has the inversion symmetry and only one of them, i.e., the projectively represented inversion operator, is permitted by the gauge condition. So we cannot determine the inversion operator of the system by simply checking whether the symmetry operator satisfies the symmetry condition. The gauge condition in Eq. \eqref{GPT} plays a guiding role in determining the inversion operator. 

We propose a Floquet engineering scheme to inter-convert the $PT$-symmetric topological phases in the presence of $\mathbb{Z}_2$ gauge degree of freedom  between $({\mathcal P}T)^2=1$ and $-1$. Note that our scheme is only applicable to systems with the $\mathbb{Z}_2$ gauge degree of freedom. A minimal requirement is that the periodic driving must not break the $PT$ symmetry. We use a two-step periodic driving protocol as
\begin{equation}\label{FE}
\begin{split}
\mathcal{H}(\textbf{k},t)= \left \{
 \begin{array}{ll}
\mathcal{H}_1(\textbf{k}),                    & t\in [n\mathscr{T},n\mathscr{T}+\mathscr{T}_{1})\\
\mathcal{H}_2(\textbf{k}),                    & t\in [n\mathscr{T}+\mathscr{T}_{1},(n+1)\mathscr{T})
 \end{array}
 \right.,
 \end{split}
 \end{equation}
where $n \in \mathbb{Z}$ and $\mathscr{T}=\mathscr{T}_1+\mathscr{T}_2$ is the driving period. The time-periodic $\mathcal{H}(\textbf{k},t)$ does not have a well-defined energy spectrum. According to the Floquet theorem, the one-period evolution operator $U_\mathscr{T}=e^{-i\mathcal{H}_{2}(\textbf{k})\mathscr{T}_{2}}e^{-i\mathcal{H}_{1}(\textbf{k})\mathscr{T}_{1}}$ defines an effective Hamiltonian $\mathcal{H}_{\text{eff}}(\textbf{k})=\frac{i}{\mathscr{T}}\ln{ U_\mathscr{T}}$, whose eigenvalues are called quasienergies. The topological features of our periodic system are defined in the quasienergy spectrum. However, $\mathcal{H}_{\textrm{eff}}({\bf k})$ does not inherit the time-reversal and chiral symmetries under the operations $T$ and $S$ of the static system due to $[\mathcal{H}_{1}(\textbf{k}), \mathcal{H}_{2}(\textbf{k})]\neq 0$. To define the time-reversal and chiral operators for the periodic system, we perform the transformations $F_{l}(\textbf{k})=e^{i(-1)^{l}\mathcal{H}_{l}(\textbf{k})\mathscr{T}_{l}/2}$ ($l=1,2$) on $U_\mathscr{T}$ and obtain \cite{PhysRevB.90.125143} 
\begin{eqnarray}
U_{\mathscr{T},1} (\textbf{k})&=&e^{-i\mathcal{H}_{1}(\textbf{k})\mathscr{T}_{1}/2}e^{-i\mathcal{H}_{2}(\textbf{k})\mathscr{T}_{2}}e^{-i\mathcal{H}_{1}(\textbf{k})\mathscr{T}_{1}/2},\label{u1}\\
U_{\mathscr{T},2} (\textbf{k})&=&e^{-i\mathcal{H}_{2}(\textbf{k})\mathscr{T}_{2}/2}e^{-i\mathcal{H}_{1}(\textbf{k})\mathscr{T}_{1}}e^{-i\mathcal{H}_{2}(\textbf{k})\mathscr{T}_{2}/2}.    \label{u2}
\end{eqnarray}
The use of $T\mathcal{H}_l({\bf k})T^{-1}=\mathcal{H}_l(-{\bf k})$ and $S\mathcal{H}_l({\bf k})S^{-1}=-\mathcal{H}_l({\bf k})$ readily leads to \cite{PhysRevB.96.195303}
\begin{eqnarray}
    T U_{\mathscr{T},l} (\textbf{k})T^{-1}&=&U_{\mathscr{T},l}^{\dag} (-\textbf{k}),\\
    S U_{\mathscr{T},l} (\textbf{k})S^{-1}&=&U_{\mathscr{T},l}^{-1} (\textbf{k}).\label{dccf}
\end{eqnarray}
Hence, the transformations restore the time-reversal and chiral symmetries on $U_{\mathscr{T},l} (\textbf{k})$ under the same operations $T$ and $S$ as those in the static system. It implies that, after making the inverse transformations $F_{l}^{-1}(\textbf{k})$, $\mathcal{H}_{\text{eff}}({\bf k})$ defined in $U_{\mathscr{T}}({\bf k})$ possesses the time-reversal and chiral symmetries under the redefined operations $\mathcal{T}_l=F_{l}^{-1}(-\textbf{k})TF_{l}(\textbf{k})$ and $\mathcal{S}_l=F_{l}^{-1}(\textbf{k})SF_{l}(\textbf{k})$. Since the internal symmetries of the static system are inherited by our periodically driven system, it is natural to expect that the periodic driving could not generate exotic topological phases according to the conventional rule of the Altland-Zirnbauer symmetry classification \cite{RevModPhys.88.035005}.  

It is surprising to find that the gauge condition \eqref{GPT} would be broken by the periodic driving that satisfies $[G,\mathcal{T}_l] \neq 0$ due to the noncommutativity between $G$ and $\mathcal{H}_{l}(\textbf{k})$ in $\mathcal{T}_l$. It causes the $PT$ symmetry class in our periodic system to go back to the original case, i.e., $(PT)^2=1$ in spinless system and $(PT)^2=-1$ in spin-$1/2$ system. This gives an idea to convert the $PT$-symmetric topological phases from $(PT)^2=\mp1$ to $\pm 1$ via manipulating the gauge degree of freedom by Floquet engineering. The underlying physical picture of this method, which uses periodic driving to convert $PT$-symmetric classes, is that the non-commutativity of the effective Hamiltonian under a step-like driving protocol twists the gauge field induced by the flux. In contrast to the widely used scheme to induce topological phase transitions between different Altland-Zirnbauer symmetry classes by changing the internal symmetries \cite{PhysRevB.103.115308,PhysRevX.3.031005,RevModPhys.89.011004}, our scheme opens another avenue to explore exotic topological phases by changing the $PT$ symmetry class via the $\mathbb{Z}_2$ gauge degree of freedom induced by the $\pi$ phase difference between different hopping rates. From an application perspective, it lays a foundation for designing quantum devices based on such conveniently controllable phases. Our scheme is applicable in systems with the $\mathbb{Z}_2$ gauge degree of freedom by breaking the $\mathbb{Z}_2$ gauge degree of freedom through Floquet engineering to realize the conversion of the $PT$-symmetric topological classes. This is in sharp contrast to Ref. \cite{PhysRevLett.126.196402}, where the interconversion of different topological classes was realized by introducing the $\mathbb{Z}_2$ gauge degree of freedom into gauge-free systems. 

\section{Extended Benalcazar-Bernevig-Hughes model}
\begin{figure}[tbp]
\centering
\includegraphics[width=\columnwidth]{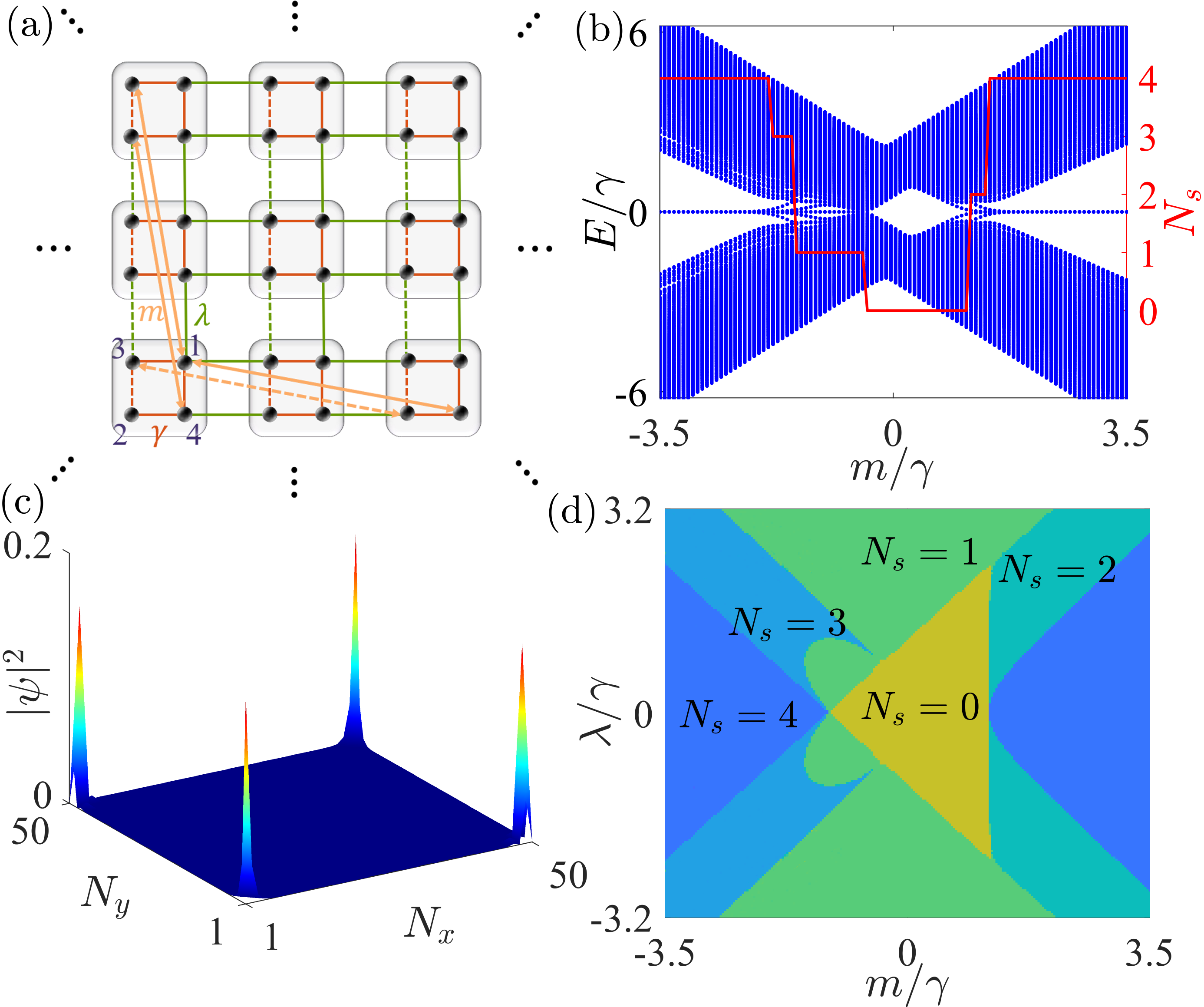}
\caption{(a) Scheme of our system. $\gamma$ and $\lambda$ are the intracell and nearest-neighbor intercell hopping rates, and $m$ is the long-range one. The dashed lines denote the hopping rates with a $\pi$-phase difference from their solid counterparts. (b) Energy spectrum and chiral number $N_s$ as a function of $m$, with $\lambda = 0.64\gamma$. (c) Distribution of the zero-energy state. (d) Phase diagram described by $N_s$. }\label{fig:2}
\end{figure}

To validate our approach and realize the inter-conversion of the $PT$-symmetric topological classes, we should choose a suitable system with $\mathbb{Z}_2$ gauge degree of freedom. Based on this requirement, we choose the Benalcazar-Bernevig-Hughes model with long-range hoppings, which is a two-dimensional spinless fermion square lattice system, see Fig. \ref{fig:2}(a). Its momentum-space Hamiltonian reads ${H}=\sum_{\textbf{k}}{C}^{\dagger}_{\textbf{k}}\mathcal{H}(\textbf{k}){C}_{\textbf{k}}$ with ${C}^{\dagger}_{\textbf{k}} = ({C}^{\dagger}_{\textbf{k},1}$ ${C}^{\dagger}_{\textbf{k},2}$ ${C}^{\dagger}_{\textbf{k},3}$ ${C}^{\dagger}_{\textbf{k},4}$) and \cite{PhysRevLett.128.127601}
\begin{equation} \label{BBH}
\mathcal{H}(\textbf{k})=\mathbf{d}(\textbf{k})\cdot\mathbf{\Gamma},
\end{equation}
where $d_1(\textbf{k})=-(\lambda\sin k_{y}+m\sin 2k_x)$, $d_2(\textbf{k})=-(\gamma+\lambda\cos k_{y}+m\cos 2k_{x})$, $d_3(\textbf{k})=-(\lambda\sin k_{x}+m\sin 2k_y)$, $d_4(\textbf{k})=0$, and $d_5(\textbf{k})=\gamma+\lambda\cos k_{x}+m\cos 2k_{y}$. $\gamma$ is the intarcell hopping rate, $\lambda$ is the intercell hopping rate, $m$ is the long-range hopping rate, $\Gamma_{i}=\tau_{y}\sigma_{i}$ ($i=1,2,3$), $\Gamma_{4}=\tau_{z}\sigma_{0}$, and $\Gamma_{5}=\tau_{x}\sigma_{0}$, with $\tau_{i}$ and $\sigma_{i}$ being Pauli matrices, $\tau_{0}$ and $\sigma_{0}$ being identity matrices. The system possesses the internal particle-hole ${\mathcal{C}=\tau_z\sigma_0K}$, time-reversal ${T=K}$, with $K$ being the complex conjugation, and chiral $S=\tau_z\sigma_0$ symmetries. According to the Altland-Zirnbauer classification rule, our system belongs to the BDI class and is topologically trivial in first order \cite{RevModPhys.88.035005}. Its external inversion operation is $P=\tau_0\sigma_x$, which cannot make the system invariant. Because of the presence of the $\pi$ phase difference between different hopping rates, the system has an additional $\mathbb{Z}_2$ gauge degree of freedom and the gauge transformation $G=\tau_0\sigma_z$ satisfies Eq. \eqref{GPT}. Therefore, the operation under which the $PT$ symmetry is respected is $\mathcal{P}T=GPT=i\tau_0\sigma_yK$, which obeys $(\mathcal{P}T)^2=-1$. According to the classification rule of the $PT$-symmetric topological phases, our system does not have real Chern insulator and only hosts the second-order topological phases \cite{PhysRevLett.116.156402,PhysRevLett.126.196402}. The chiral symmetry makes the real-space Hamiltonian $H$ of this system unitarily equivalent to $H'=\left(
  \begin{array}{cc}
    0 & D \\
    D^\dag & 0 \\
  \end{array}
\right)$. 
Thus, the lattice in the real space is split into two sublattices labeled by $A$ and $B$. Its second-order topological phase is described by the chiral number \cite{PhysRevLett.128.127601}
\begin{equation}\label{NS}
N_s=\frac{1}{2\pi i}\textrm{Tr}\, \textrm{log}(\mathcal{Q}_A\mathcal{Q}_B^\dagger),
\end{equation}
where $\mathcal{Q}_p=\mathcal{U}_p^\dagger Q_p\mathcal{U}_p$ and $Q_p=\sum_{\textbf{R},\beta \in p}c^\dagger_{\textbf{R},\beta}|0\rangle \exp\big(-i\frac{2\pi xy}{L_xL_y}\big)\langle 0|c_{\textbf{R},\beta}$ are the sublattice quadrupole moment operators, with $p=A,B$ and $c^\dagger_{\textbf{R},\beta}$ being the creation operator of the fermion in the $p$ sublattice of the unit cell $\textbf{R}=(x,y)$. The unitary transformation $\mathcal{U}_p=(\psi_1^p, \psi_2^p,\cdots, \psi_{M}^p)$, with $DD^\dagger\psi_n^A=\epsilon_n^2\psi_n^A$, $D^\dagger D\psi_n^B=\epsilon_n^2\psi_n^B$, and $M=2L_xL_y$, is performed to project $Q_p$ in the space spanned by $\{\psi_n^p\}$. Note that chiral symmetry is just for the convenience in defining the chiral number. It is not necessary for our result to switch different topological classes. The energy spectrum under the open-boundary condition in Fig. \ref{fig:2}(b) reveals that $4|N_s|$ degenerate zero-energy states are formed. They are fully separated with finite gaps from the bulk energies. Their probability distribution in Fig. \ref{fig:2}(c) confirms the second-order corner-state nature. We call these states the gapped corner states. The widely tunable $N_s$ with the change of $m$ originates from the long-range hoppings of our model. The phase diagram in Fig. \ref{fig:2}(d) gives a global picture of the topological phases. Our result verifies that the system only hosts the second-order topological phases. It agrees with the classification rule of the $PT$-symmetric topological phase with $(\mathcal{P}T)^2=-1$.

\begin{figure}[tbp]
\centering
\includegraphics[width=\columnwidth]{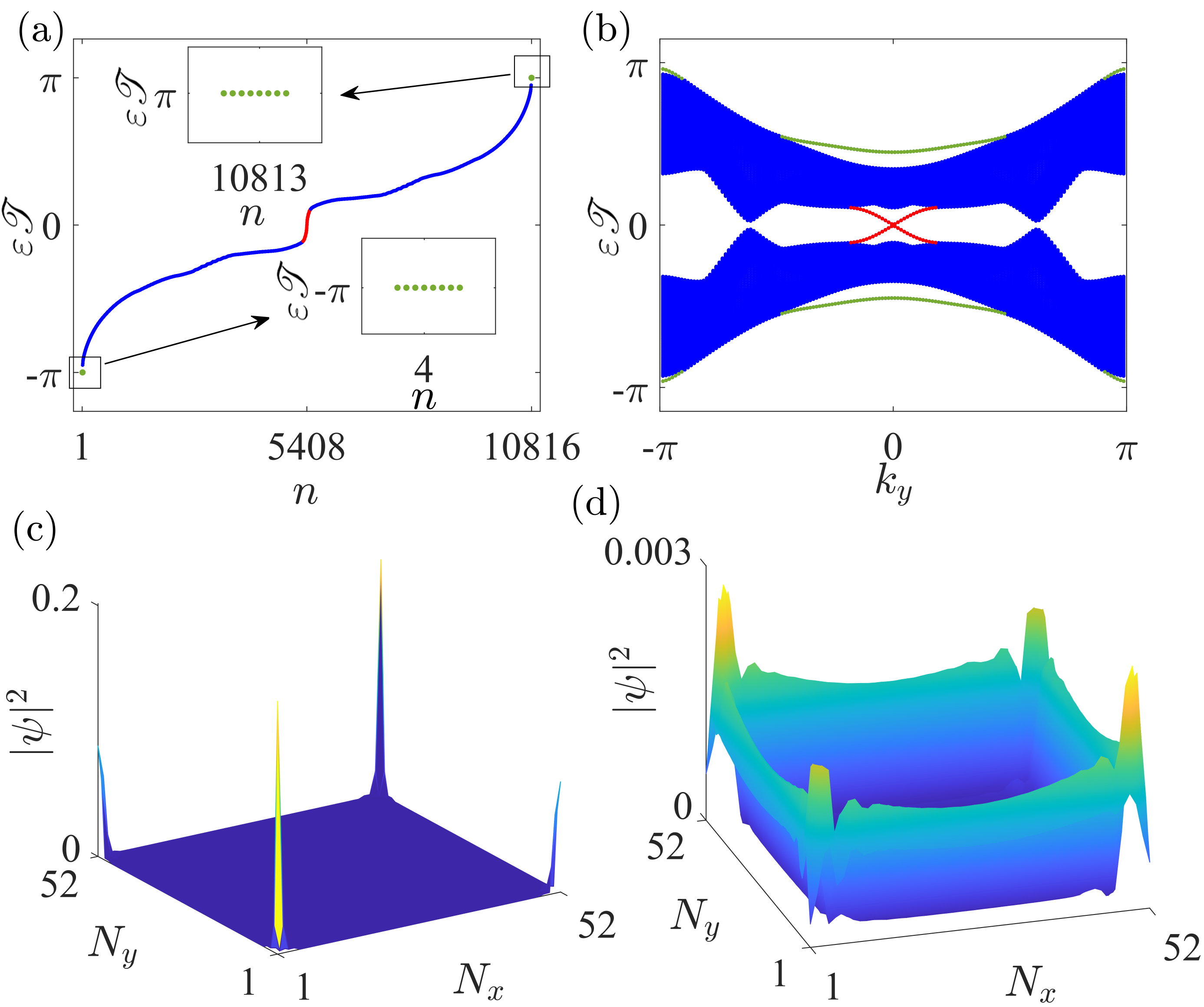}
\caption{Quasienergies under the open boundary conditions of (a) the $x$ and $y$ directions and (b) only the $x$ direction. The red line in (a) and (b) represent the gapless boundary states. The green point in (a) and line in (b) represent the gapped corner states. Distributions of (c) the gapped $|\pi/\mathscr{T}|$-quasienergy state and (d) the gapless zero-quasienergy state. We use $\lambda=0.64\gamma$, $m_1=-0.36\gamma$, $m_2=3.6\gamma$, $\mathscr{T}_1=1.2\gamma^{-1}$ and $\mathscr{T}_2=0.6\gamma^{-1}$. }\label{fig:3}
\end{figure}

To generate rich $PT$-symmetric topological phases, we intend to convert the symmetry class into the one with $(\mathcal{P}T)^2=1$. We apply a periodic driving on the hopping rate $m$ as 
\begin{equation}\label{FE1}
\begin{split}
m(t)= \left \{
 \begin{array}{ll}
m_1,                    & t\in [n\mathscr{T},n\mathscr{T}+\mathscr{T}_{1})\\
m_2,                    & t\in [n\mathscr{T}+\mathscr{T}_{1},(n+1)\mathscr{T})
 \end{array}
 \right..
 \end{split}
 \end{equation}
According to Eq. \eqref{dccf}, $\mathcal{H}_{\text{eff}}(\textbf{k})$ has the time-reversal and chiral symmetries under the redefined operations $\mathcal{T}_l$ and $\mathcal{S}_l$. We can prove $[G,\mathcal{T}_l]\neq 0$. Explicitly, one of the redefined time-reversal operators is $\mathcal{T}_1=e^{i\mathcal{H}_1(-{\bf k})\mathscr{T}_1/2}Te^{-i\mathcal{H}_1({\bf k})\mathscr{T}_1/2}$, with $\mathcal{H}_1({\bf k})=\mathbf{d}_1\cdot{\pmb \Gamma}$. It is easy to verify $ [\Gamma_{3/5},G]=0$ and $[\Gamma_{1/2},G]\neq 0$. Therefore, we have $[\mathcal{T}_1,G]\neq 0$ in the periodically driven case even $[T,G]=0$ in the static case. The same conclusion can be drawn from $\mathcal{T}_2$. Thus, the prerequisite Eq. \eqref{GPT} of the $\mathbb{Z}_2$ gauge degree of freedom is invalidated and $\mathcal{P}=GP$ cannot describe the inversion operation of the periodic system anymore. The space-time inversion symmetry of the periodically driven system goes back to the one of the original case $(P'\mathcal{T}_l)^2=1$ in spinless system without $\mathbb{Z}_2$ gauge degree of freedom, which results in $P'=\tau_{0}\sigma_{0}I$, with $I$ being the momentum inversion operation \cite{PhysRevLett.116.156402,PhysRevLett.126.196402,PhysRevLett.118.056401}. Therefore, we realize the conversion of the $PT$-symmetric topological phases from $(\mathcal{P}T)^2=-1$ to $1$ via breaking the gauge degree of freedom by periodic driving. This is impossible in static systems. It is emphasized that our periodic driving neither breaks the $\mathcal{P}T$ symmetry and nor creates any new symmetry. In our periodic and static systems, both $\mathcal{P}$ and $P'$ satisfy the inversion-symmetry condition. In the static system, the $\mathbb{Z}_2$ gauge condition is met. Therefore, $\mathcal{P}T=GPT$ is the correct symmetry operator. In the periodically driven system, the inversion operator cannot be projectively represented as $\mathcal{P}$ because the $\mathbb{Z}_2$ gauge condition is no longer met. In this case, the correct inversion operator becomes $P'$. The conventional Altland-Zirnbauer topological classification of the system does not change because the three internal symmetries are fully inherited in our periodic system. This endows our work with a substantial difference from the traditional Floquet topological states, where periodic driving induces exotic topological phases mainly by changing the internal symmetries of the systems \cite{McIver2020,PhysRevB.103.045424,PhysRevLett.121.076802,PhysRevLett.117.087402,PhysRevB.105.L081102}. 
\begin{figure}[tbp]
\centering
\includegraphics[width=\columnwidth]{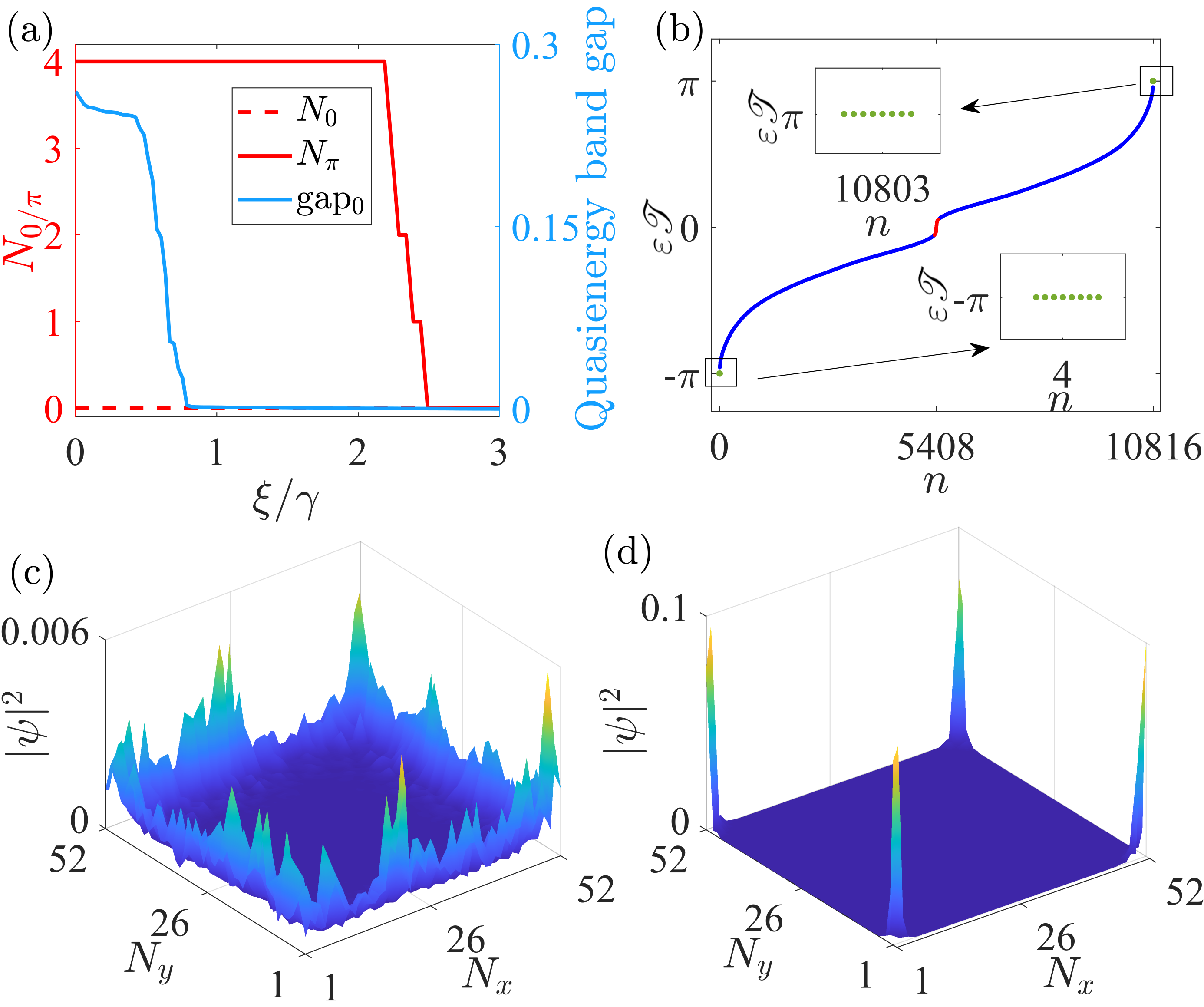}
\caption{(a) Chiral numbers and zero-mode quasienergy band gap with different disorder strengths $\xi$. (b) Quasienergies under the open boundary conditions of the $x$ and $y$ directions with disorder strengths $\xi=0.5$. The green point and red line represent the gapped corner states and gapless boundary states, respectively. Distribution of (c) the zero-mode boundary state and (d) $|\pi/\mathscr{T}|$-mode corner state with disorder strengths $\xi=0.5\gamma$. We use $\lambda=0.64\gamma$, $m_1=-0.36\gamma$, $m_2=3.6\gamma$, $\mathscr{T}_1=1.2\gamma^{-1}$ and $\mathscr{T}_2=0.6\gamma^{-1}$.}\label{fig:4}
\end{figure}

According to the classification rule of the $PT$-symmetric topological phases \cite{PhysRevLett.116.156402,PhysRevLett.126.196402}, our periodic system with $(P\mathcal{T}_l)^2=1$ supports the first-order real Chern insulators and the second-order topological insulators. It is noted that the systems with $(PT)^2=-1$ do not have first-order topological insulators \cite{PhysRevLett.116.156402,PhysRevLett.126.196402}. So, the emergence of the first-order topological insulators in our periodic system is associated with conversion of different $PT$-symmetric classes. And the systems with $(PT)^2=1$ or $(PT)^2=-1$ all host the second-order topological insulators \cite{PhysRevLett.116.156402,PhysRevLett.126.196402}. So the second-order topological insulators are protected by chiral symmetry. Furthermore, the topological phases of the periodic system occur not only at quasienergy zero but also at $|\pi/\mathscr{T}|$. Thus, setting the $PT$-symmetric topological phases free from the constraint of $(P\mathcal{T}_l)^2=-1$, extremely rich topological phases emerge in our periodic system. Now, we develop a complete characterization of them. First, we select two different initial times, $t_1$ and $t_2$, to define two one-periodic evolution operators. We ensure that their zero-quasienergy quasistationary states are on the same chirality sublattice, while their $\pi/\mathscr{T}$-quasienergy quasistationary states are on the opposite chirality sublattice \cite{PhysRevB.90.125143}. 
\begin{figure}[tbp]
\centering
\includegraphics[width=\columnwidth]{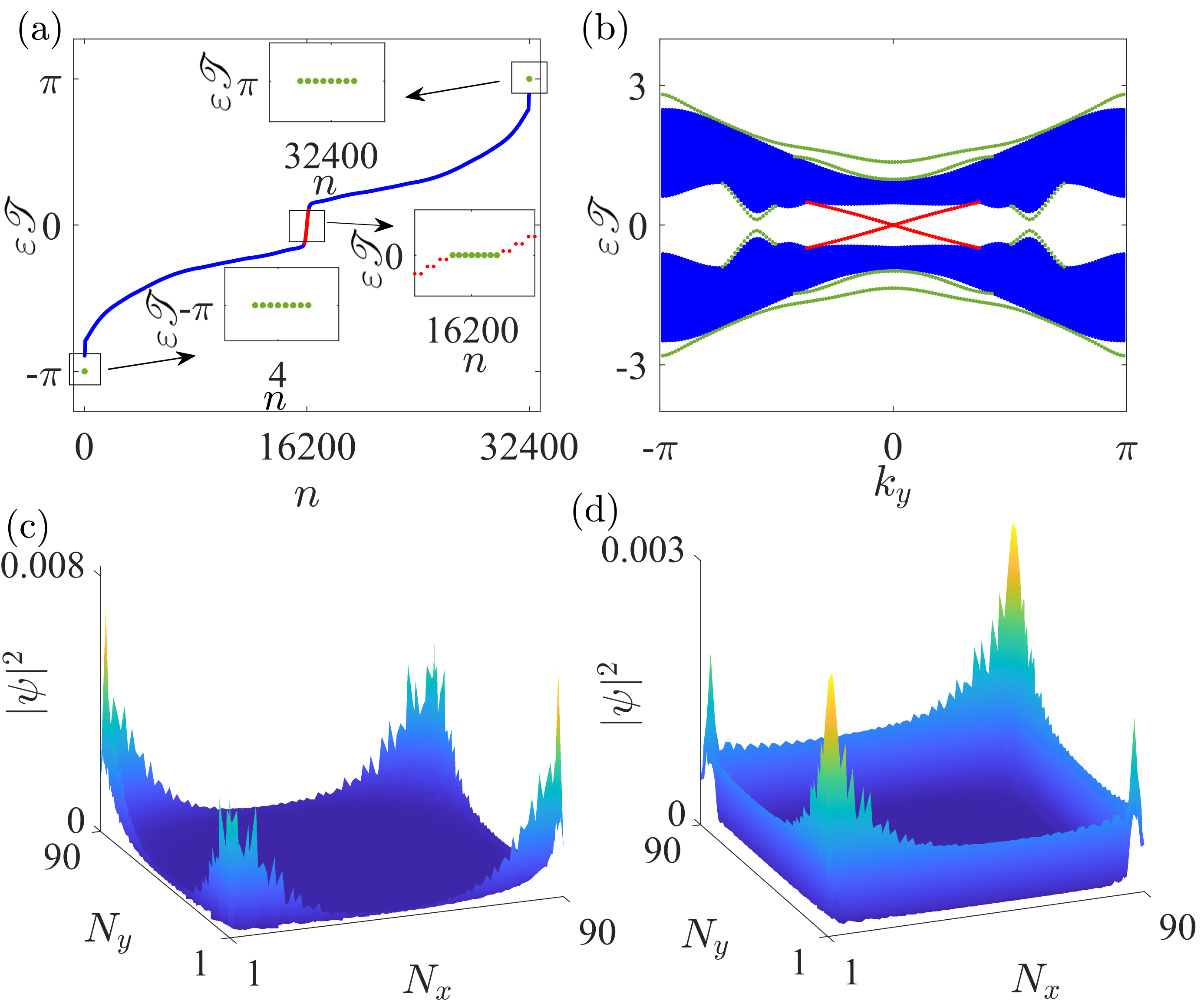}
\caption{(a) Quasienergies under the open boundary conditions of (a) the $x$ and $y$ directions and (b) only the $x$ direction. The red line in (a) and (b) represent the gapless boundary states. The green point in (a) and line in (b) represent the gapped corner states. Distribution of (c) the zero-mode corner state and (d) boundary state. We use $\lambda=0.64\gamma$, $m_1=-0.1\gamma$, $m_2=3.6\gamma$, $\mathscr{T}_1=1.2\gamma^{-1}$ and $\mathscr{T}_2=0.6\gamma^{-1}$. }\label{fig:5}
\end{figure}
Thus, making two unitary transformations $F_l({\bf k})$ on $U_\mathscr{T}({\bf k})$, we can define two Hamiltonians from the transformed evolution operators $U_{\mathscr{T},l}({\bf k})$ as $\mathbb{H}_{\text{eff},l}(\textbf{k})=i\mathscr{T}^{-1}\ln[{U}_{\mathscr{T},l}(\textbf{k})]$, which have the same quasienergy spectrum as $\mathcal{H}_\text{eff}({\bf k})$. ${U}_{\mathscr{T},1}(\textbf{k})$ and ${U}_{\mathscr{T},2}(\textbf{k})$ correspond to the two initial times $t_1=\mathscr{T}_1/2$ and $t_2=\mathscr{T}_1+\mathscr{T}_2/2$ and satisfy the aforementioned requirements. In a similar manner as in Eq. \eqref{NS} in the static system, we define two chiral numbers $N_{s,l}$ in $\mathbb{H}_{\text{eff},l}(\textbf{k})$. The second-order topological phases of the periodic system in the zero and $|\pi/\mathscr{T}|$ modes are described by $N_{\alpha}=(N_{s,1}+e^{i\alpha}N_{s,2})/2$, with $\alpha=0$ and $\pi$ \cite{PhysRevB.103.L041115,PhysRevB.104.205117,PhysRevA.100.023622,PhysRevB.102.041119,PhysRevB.107.035128,PhysRevB.109.184518,PhysRevB.110.125427}. Second, the first-order topological phases of this $PT$-symmetric system are characterized by the real Chern number \cite{PhysRevLett.118.056401,PhysRevLett.125.126403} 
\begin{equation}\label{RC}
\mathcal{V}_R=\frac{-i}{4\pi}\int_{\text{BZ}}d^2\textbf{k}\text{Tr}[\tau_{y}\sigma_{0}(\nabla_{\textbf{k}}\times \mathcal{A})_z] \quad \text{mod} \ 2,
\end{equation}
where $\mathcal{A}_{mn}= \langle m,\textbf{k}|\nabla_{\textbf{k}}|n,\textbf{k}\rangle$ and $|m/n,\textbf{k}\rangle$ are the real eigenstates of $\mathcal{H}_{\text{eff}}({\bf k})$ under the reality requirement $P\mathcal{T}_l|m/n,\textbf{k}\rangle=|m/n,\textbf{k}\rangle$. 

Figure \ref{fig:3}(a) shows the quasienergies of the periodic system under the open boundary condition. We can clearly find that sixteen gapped corner states appear in the $|\pi/\mathscr{T}|$ modes and a pair of gapless boundary states appear in the zero mode. The quasienergies in Fig. \ref{fig:3}(b) under the $x$-direction open boundary condition prove that the first-order gapless boundary states occur in the zero mode and the second-order gapped corner states occur in the $|\pi/\mathscr{T}|$ mode. The $\pi/\mathscr{T}$-mode corner states become a dispersion-relation curve with $k_y$, see the green line in Fig. \ref{fig:3}(b). It is generally believed that the second-order gapped corner states are generated by opening a gap of the first-order gapless boundary states \cite{PhysRevB.107.235132,PhysRevB.102.125144}. We really see from the green curve in Fig. \ref{fig:3}(b) that the $k_y$ dispersion relation has a finite band gap near the quasienergy $\pi/\mathscr{T}$. 
\begin{figure}[tbp]
\centering
\includegraphics[width=\columnwidth]{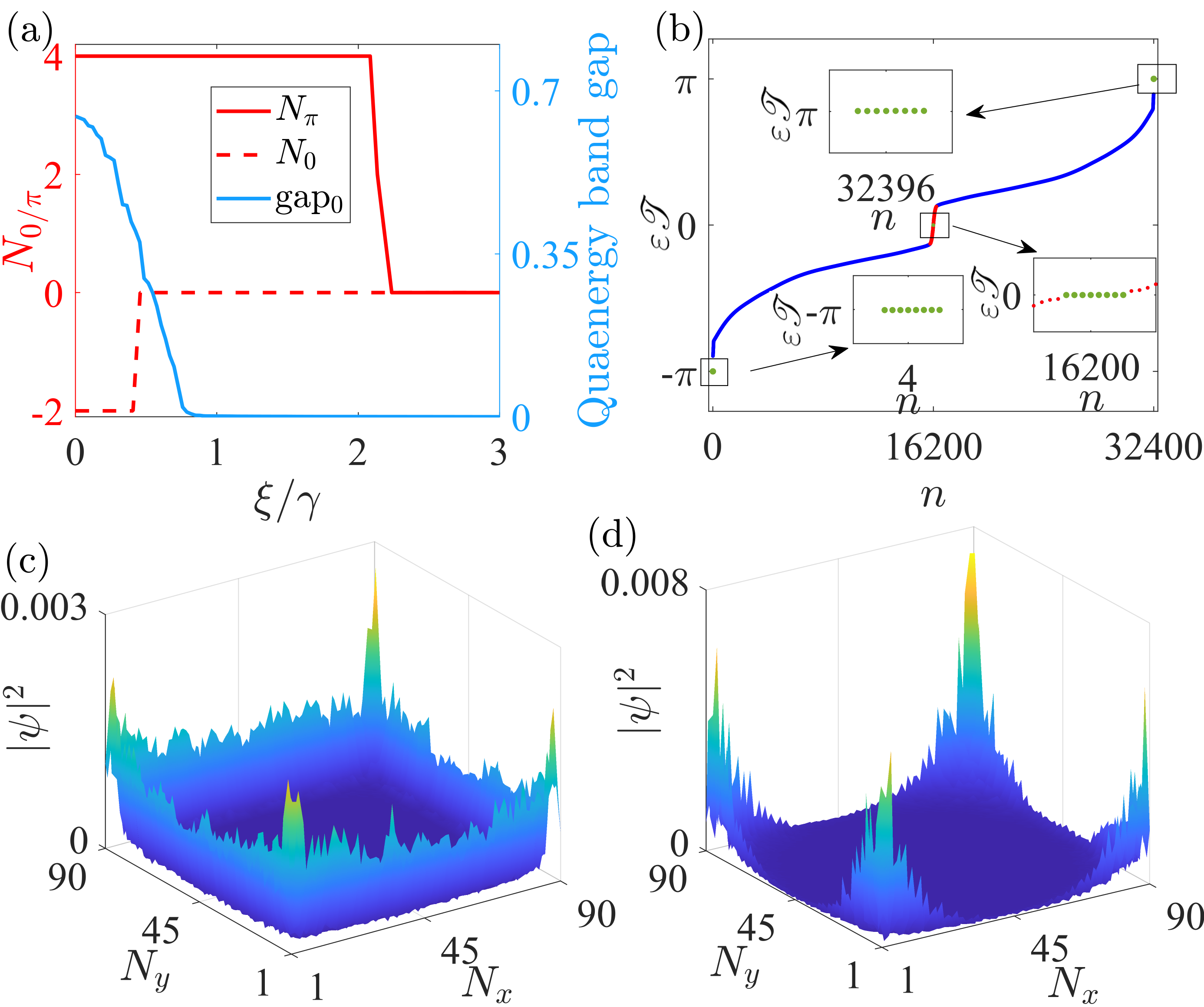}
\caption{(a) Chiral numbers and zero-mode quasienergy band gap with different disorder strengths $\xi$. (b) Quasienergies under the open boundary conditions of the $x$ and $y$ directions with disorder strengths $\xi=0.2$. The green point and red line represent the gapped corner states and gapless boundary states, respectively. Distribution of (c) the zero-mode boundary state and (d) corner state with disorder strengths $\xi=0.2\gamma$. We use $\lambda=0.64\gamma$, $m_1=-0.1\gamma$, $m_2=3.6\gamma$, $\mathscr{T}_1=1.2\gamma^{-1}$ and $\mathscr{T}_2=0.6\gamma^{-1}$. }\label{fig:6}
\end{figure}
It is a reflection of the quasienergy spectrum of Fig. \ref{fig:3}(a) under a different boundary condition. This is further confirmed by the probability distribution of the $\pi/\mathscr{T}$- and zero-gap states in Fig. \ref{fig:3}(c) and \ref{fig:3}(d). Calculating the chiral numbers and the real Chern number, we obtain $N_{0}=0$, $N_{\pi}=4$, and $\mathcal{V}_R=1$. They quantify $4|N_{\pi}|$ second-order corner states in the $|\pi/\mathscr{T}|$ mode and $\mathcal{V}_R$ pair of first-order boundary states in the zero mode. It should be emphasized that this result is not a finite-size effect. Since quasienergy spectrum in Fig. \ref{fig:3}(b) is obtained under the periodic boundary condition of $y$ direction, it is independent of the system size of $N_y$. Defined in the bulk bands under the periodic boundary condition along $x$ and $y$ directions, all the topological invariants are independent of the system sizes $N_{x}$ and $N_y$ too. To verify the topological protection of corner and boundary states, we consider a disorder $\xi\eta$ on the hopping rate $m_1$ and $m_2$, where $\eta \in[-1,1]$ is a random number and $\xi$ is the disorder strength. The hopping rate has become $m_1+\xi\eta$ and $m_2+\xi\eta$. This disorder breaks the $P'\mathcal{T}_l$ symmetry, but preserves chiral symmetry. Therefore, we can still use chiral numbers to characterize the topological properties of second-order topological phase. The topological properties of first-order topological phase can be characterized by zero-mode quasienergy band gap. Figure \ref{fig:4}(a) shows the topological phases in our periodic system are robust to the weak disorder. And the quasienergies and the probability distributions of the zero-mode boundary state and $|\pi/\mathscr{T}|$-mode corner state when the disorder strength $\xi=0.5\gamma$ also demonstrate that the topological phase has not been broken, see Fig. \ref{fig:4}(b)-(d). Thus, we realize the coexistence of the first-order real Chern insulator in zero mode and the second-order topological insulator in $|\pi/\mathscr{T}|$ mode in a $PT$-symmetric system via converting its topological class from $(\mathcal{P}T)^2=-1$ to $1$ by the periodic driving. 

Next, the periodic driving also makes the first- and second-order topological phases coexist even in the same quasienergy gap. Figure \ref{fig:5}(a) shows the quasienergies under the open boundary condition of the $x$ and $y$ directions. It can be seen that, in addition to the gapped $|\pi/\mathscr{T}|$ mode, the corner states are also present in the quasienergy band of the gapless boundary states of the zero mode. This is confirmed by the quasienergies under the $x$-direction open boundary condition in Fig. \ref{fig:5}(b). The probability distribution of the state with quasienergy zero in Fig. \ref{fig:5}(c) shows the feature of the second-order corner mode. The one of the states with quasienergy nearest to zero in Fig. \ref{fig:5}(d) clearly shows the feature of the first-order boundary mode. They verify the coexistence of the boundary and corner states at the zero gap. The topological invariants are calculated as $N_{0}=-2$, $N_{\pi}=4$, and $\mathcal{V}_R=1$. They indicate that eight second-order corner states and a pair of first-order gapless boundary states coexist in the zero gap and sixteen second-order corner states exist in the $|\pi/\mathscr{T}|$ gap. It is noted that our result is also not a finite-size effect. The reason is the same as in the previous ones. Since quasienergy spectrum in Fig. \ref{fig:5}(b) is independent of the system size of $N_y$ and all the topological invariants are independent of the system sizes $N_{x}$ and $N_y$. Similarly, we can use a disorder that is same to the previous one to verify the topological protection of this novel topological phase. Figure \ref{fig:6}(a) also demonstrates the robustness of this topological phase. And the quasienergies and the probability distributions of the zero-mode boundary state and corner state when the disorder strength $\xi=0.2\gamma$ demonstrate that the topological phase has not been broken, see Fig. \ref{fig:6}(b)-(d). Thus, we achieve the coexistence of the first- and second-order topological insulators in the same gap by Floquet engineering. Such an exotic $PT$-symmetric topological phase cannot be realized in static systems. The coexisting first- and second-order topological insulators in a periodically driven $PT$-symmetric system was reported in Ref. \cite{xz4g-jlbg}. However, they reside in different quasienergy gaps. The coexisting first- and second-order topological insulators in one single quasienergy gap have not been reported. It is generally believed that when distinct boundary and corner states reside at the same quasienergy gap in a finite-sized lattice, they typically have some spatial overlap and tend to hybridize. However, the first-order topological phase described by the real Chern number in our periodic system is protected by $PT$ symmetry \cite{PhysRevLett.118.056401}. The second-order topological phase described by the chiral number in our periodic system is protected by chiral symmetry \cite{PhysRevLett.128.127601}. Thus, in our periodic system, the second- and first-order topological phases can coexist in same quasienergy gap without hybridization or wavefunction overlap.

\section{Discussion and conclusion}
It should be noted that although only the step-like driving protocol is considered, our scheme is generalizable to other driving protocols. It should be noted that the real Chern number Eq. \eqref{RC} is defined by all occupied bands. It simultaneously contains information about both the zero and $\pi/\mathscr{T}$ quasienergy gaps simultaneously. Therefore, we can only characterize the case where the first-order boundary modes exist either in the zero quasienergy gap or the $\pi/\mathscr{T}$ quasienergy gap. If first-order boundary modes simultaneously appear in both the zero and $\pi/\mathscr{T}$ quasienergy gaps, we currently have no means of characterizing them. Here, our primary result demonstrates that periodic driving can convert $PT$-symmetric classes. Thus, this insufficiency does not affect our main conclusion. The $PT$-symmetric topological phases have been realized in photonic, acoustic, and electronic systems \cite{PhysRevLett.128.116803,PhysRevLett.130.026101,PhysRevLett.132.197202,PhysRevLett.133.176602,Pan2023}. And the two-dimensional extended Benalcazar-Bernevig-Hughes model is usually implemented in acoustic crystal \cite{PhysRevLett.131.157201,PhysRevB.108.205135} and electrical circuits \cite{PhysRevApplied.20.064042}. Floquet engineering has become a versatile tool for generating novel topological phases on various platforms \cite{McIver2020,Wintersperger2020,Maczewsky2017,PhysRevLett.122.173901,PhysRevLett.129.254301,PhysRevLett.133.073803}. These progresses give strong support to the experimental realization of our scheme. 

In summary, we have proposed a scheme to convert the $PT$-symmetric topological phases from $(\mathcal{P}T)^2=-1$ to $1$ by Floquet engineering. Breaking through the constraint of $(\mathcal{P}T)^2=-1$, we have discovered diverse exotic $PT$-symmetric phases with coexisting first- and second-order topological insulators not only in different quasienergy gaps but also in one single quasienergy gap. They are completely absent in the corresponding static system. Our study reveals that it is the removal of the $\mathbb{Z}_2$ gauge degree of freedom by the periodic driving that causes the inter-conversion between different symmetry classes. Without changing any symmetry, our scheme has a substantial difference from the conventional Floquet engineering, which pursues exotic topological phases by breaking the internal symmetries. This refreshes our general belief of Floquet topological phases and broadens the frontier. Enriching the family of $PT$-symmetric topological phases, our result opens an avenue to controllably explore novel topological phases via removing the $\mathbb{Z}_2$ gauge degree of freedom instead of breaking the symmetries and to design multifunctional quantum devices by simultaneously utilizing the respective advantages of different orders of topological phases.

%
%

\ack{The work is supported by the National Natural Science Foundation of China (Grants No. 124B2090, No. 12275109, and No. 12247101), the Quantum Science and Technology-National Science and Technology Major Project (Grant No. 2023ZD0300904), the Fundamental Research Funds for the Central Universities (Grant No. lzujbky-2024-jdzx06), and the Natural Science Foundation of Gansu Province (Grant No. 22JR5RA389).}

\roles{Jun-Hong An proposed the original idea and developed the theoretical formalism. Ming-Jian Gao performed the analytic derivations and drafted the manuscript and designed the figures. All authors discussed the results and contributed to the final manuscript.}

\data{All data that support the findings of this study are included within the article. They are available from the corresponding author upon reasonable request.}

\bibliographystyle{iopart-num}
\bibliography{references}

\providecommand{\newblock}{}
\begin{thebibliography}{10}
\expandafter\ifx\csname url\endcsname\relax
  \def\url#1{{\tt #1}}\fi
\expandafter\ifx\csname urlprefix\endcsname\relax\def\urlprefix{URL }\fi
\providecommand{\eprint}[2][]{\url{#2}}

\bibitem{RevModPhys.82.3045}
Hasan M~Z and Kane C~L 2010 {\em Rev. Mod. Phys.\/} {\bf 82}(4) 3045--3067
  \urlprefix\url{https://link.aps.org/doi/10.1103/RevModPhys.82.3045}

\bibitem{RevModPhys.83.1057}
Qi X~L and Zhang S~C 2011 {\em Rev. Mod. Phys.\/} {\bf 83}(4) 1057--1110
  \urlprefix\url{https://link.aps.org/doi/10.1103/RevModPhys.83.1057}

\bibitem{RevModPhys.87.137}
Elliott S~R and Franz M 2015 {\em Rev. Mod. Phys.\/} {\bf 87}(1) 137--163
  \urlprefix\url{https://link.aps.org/doi/10.1103/RevModPhys.87.137}

\bibitem{RevModPhys.88.035005}
Chiu C~K, Teo J~C~Y, Schnyder A~P and Ryu S 2016 {\em Rev. Mod. Phys.\/} {\bf
  88}(3) 035005
  \urlprefix\url{https://link.aps.org/doi/10.1103/RevModPhys.88.035005}

\bibitem{RevModPhys.90.015001}
Armitage N~P, Mele E~J and Vishwanath A 2018 {\em Rev. Mod. Phys.\/} {\bf
  90}(1) 015001
  \urlprefix\url{https://link.aps.org/doi/10.1103/RevModPhys.90.015001}

\bibitem{RevModPhys.93.025002}
Lv B~Q, Qian T and Ding H 2021 {\em Rev. Mod. Phys.\/} {\bf 93}(2) 025002
  \urlprefix\url{https://link.aps.org/doi/10.1103/RevModPhys.93.025002}

\bibitem{RevModPhys.88.021004}
Bansil A, Lin H and Das T 2016 {\em Rev. Mod. Phys.\/} {\bf 88}(2) 021004
  \urlprefix\url{https://link.aps.org/doi/10.1103/RevModPhys.88.021004}

\bibitem{doi:10.1126/science.aah6442}
Benalcazar W~A, Bernevig B~A and Hughes T~L 2017 {\em Science\/} {\bf 357}
  61--66
  \urlprefix\url{https://www.science.org/doi/abs/10.1126/science.aah6442}

\bibitem{PhysRevLett.129.046802}
Tanaka Y, Zhang T, Uwaha M and Murakami S 2022 {\em Phys. Rev. Lett.\/} {\bf
  129}(4) 046802
  \urlprefix\url{https://link.aps.org/doi/10.1103/PhysRevLett.129.046802}

\bibitem{PhysRevLett.128.026405}
Chen C, Zeng X~T, Chen Z, Zhao Y~X, Sheng X~L and Yang S~A 2022 {\em Phys. Rev.
  Lett.\/} {\bf 128}(2) 026405
  \urlprefix\url{https://link.aps.org/doi/10.1103/PhysRevLett.128.026405}

\bibitem{PhysRevX.7.041069}
Kruthoff J, de~Boer J, van Wezel J, Kane C~L and Slager R~J 2017 {\em Phys.
  Rev. X\/} {\bf 7}(4) 041069
  \urlprefix\url{https://link.aps.org/doi/10.1103/PhysRevX.7.041069}

\bibitem{RevModPhys.91.015006}
Ozawa T, Price H~M, Amo A, Goldman N, Hafezi M, Lu L, Rechtsman M~C, Schuster
  D, Simon J, Zilberberg O and Carusotto I 2019 {\em Rev. Mod. Phys.\/} {\bf
  91}(1) 015006
  \urlprefix\url{https://link.aps.org/doi/10.1103/RevModPhys.91.015006}

\bibitem{RN237}
Guo Q, Jiang T, Zhang R~Y, Zhang L, Zhang Z~Q, Yang B, Zhang S and Chan C~T
  2021 {\em Nature\/} {\bf 594} 195--200 ISSN 1476-4687
  \urlprefix\url{https://doi.org/10.1038/s41586-021-03521-3}

\bibitem{PhysRevLett.125.053601}
\"Unal F~N, Bouhon A and Slager R~J 2020 {\em Phys. Rev. Lett.\/} {\bf 125}(5)
  053601
  \urlprefix\url{https://link.aps.org/doi/10.1103/PhysRevLett.125.053601}

\bibitem{Guo_2021}
Guo Q, Jiang T, Zhang R~Y, Zhang L, Zhang Z~Q, Yang B, Zhang S and Chan C~T
  2021 {\em Nature\/} {\bf 594} 195–200 ISSN 1476-4687
  \urlprefix\url{http://dx.doi.org/10.1038/s41586-021-03521-3}

\bibitem{Jiang2021}
Jiang B, Bouhon A, Lin Z~K, Zhou X, Hou B, Li F, Slager R~J and Jiang J~H 2021
  {\em Nature Physics\/} {\bf 17} 1239--1246 ISSN 1745-2481
  \urlprefix\url{https://doi.org/10.1038/s41567-021-01340-x}

\bibitem{Peng2022}
Peng B, Bouhon A, Monserrat B and Slager R~J 2022 {\em Nature Communications\/}
  {\bf 13} 423 ISSN 2041-1723
  \urlprefix\url{https://doi.org/10.1038/s41467-022-28046-9}

\bibitem{Bouhon_2020}
Bouhon A, Wu Q, Slager R~J, Weng H, Yazyev O~V and Bzdušek T 2020 {\em Nature
  Physics\/} {\bf 16} 1137--1143 ISSN 1745-2481
  \urlprefix\url{https://doi.org/10.1038/s41567-020-0967-9}

\bibitem{Slager_2024}
Slager R~J, Bouhon A and Ünal F~N 2024 {\em Nature Communications\/} {\bf 15}
  1144 ISSN 2041-1723
  \urlprefix\url{https://doi.org/10.1038/s41467-024-45302-2}

\bibitem{PhysRevB.102.161117}
Zhao Y~X, Huang Y~X and Yang S~A 2020 {\em Phys. Rev. B\/} {\bf 102}(16) 161117
  \urlprefix\url{https://link.aps.org/doi/10.1103/PhysRevB.102.161117}

\bibitem{PhysRevLett.116.156402}
Zhao Y~X, Schnyder A~P and Wang Z~D 2016 {\em Phys. Rev. Lett.\/} {\bf 116}(15)
  156402
  \urlprefix\url{https://link.aps.org/doi/10.1103/PhysRevLett.116.156402}

\bibitem{PhysRevLett.128.116803}
Li T, Du J, Zhang Q, Li Y, Fan X, Zhang F and Qiu C 2022 {\em Phys. Rev.
  Lett.\/} {\bf 128}(11) 116803
  \urlprefix\url{https://link.aps.org/doi/10.1103/PhysRevLett.128.116803}

\bibitem{PhysRevLett.128.116802}
Xue H, Wang Z, Huang Y~X, Cheng Z, Yu L, Foo Y~X, Zhao Y~X, Yang S~A and Zhang
  B 2022 {\em Phys. Rev. Lett.\/} {\bf 128}(11) 116802
  \urlprefix\url{https://link.aps.org/doi/10.1103/PhysRevLett.128.116802}

\bibitem{Peizhe_2016}
Tang P, Zhou Q, Xu G and Zhang S~C 2016 {\em Nature Physics\/} {\bf 12}
  1100--1104 ISSN 1745-2481 \urlprefix\url{https://doi.org/10.1038/nphys3839}

\bibitem{Qiu_2023}
Qiu J~X, Tzschaschel C, Ahn J, Gao A, Li H, Zhang X~Y, Ghosh B, Hu C, Wang Y~X,
  Liu Y~F, Bérubé D, Dinh T, Gong Z, Lien S~W, Ho S~C, Singh B, Watanabe K,
  Taniguchi T, Bell D~C, Lu H~Z, Bansil A, Lin H, Chang T~R, Zhou B~B, Ma Q,
  Vishwanath A, Ni N and Xu S~Y 2023 {\em Nature Materials\/} {\bf 22} 583--590
  ISSN 1476-4660 \urlprefix\url{https://doi.org/10.1038/s41563-023-01493-5}

\bibitem{PhysRevLett.125.126403}
Wang K, Dai J~X, Shao L~B, Yang S~A and Zhao Y~X 2020 {\em Phys. Rev. Lett.\/}
  {\bf 125}(12) 126403
  \urlprefix\url{https://link.aps.org/doi/10.1103/PhysRevLett.125.126403}

\bibitem{PhysRevX.8.031069}
Song Z, Zhang T and Fang C 2018 {\em Phys. Rev. X\/} {\bf 8}(3) 031069
  \urlprefix\url{https://link.aps.org/doi/10.1103/PhysRevX.8.031069}

\bibitem{PhysRevLett.121.106403}
Ahn J, Kim D, Kim Y and Yang B~J 2018 {\em Phys. Rev. Lett.\/} {\bf 121}(10)
  106403
  \urlprefix\url{https://link.aps.org/doi/10.1103/PhysRevLett.121.106403}

\bibitem{PhysRevLett.121.036401}
Li L, Lee C~H and Gong J 2018 {\em Phys. Rev. Lett.\/} {\bf 121}(3) 036401
  \urlprefix\url{https://link.aps.org/doi/10.1103/PhysRevLett.121.036401}

\bibitem{PhysRevLett.123.016806}
Peng Y and Refael G 2019 {\em Phys. Rev. Lett.\/} {\bf 123}(1) 016806
  \urlprefix\url{https://link.aps.org/doi/10.1103/PhysRevLett.123.016806}

\bibitem{PhysRevLett.124.057001}
Hu H, Huang B, Zhao E and Liu W~V 2020 {\em Phys. Rev. Lett.\/} {\bf 124}(5)
  057001
  \urlprefix\url{https://link.aps.org/doi/10.1103/PhysRevLett.124.057001}

\bibitem{PhysRevLett.124.216601}
Huang B and Liu W~V 2020 {\em Phys. Rev. Lett.\/} {\bf 124}(21) 216601
  \urlprefix\url{https://link.aps.org/doi/10.1103/PhysRevLett.124.216601}

\bibitem{PhysRevLett.117.087402}
Yan Z and Wang Z 2016 {\em Phys. Rev. Lett.\/} {\bf 117}(8) 087402
  \urlprefix\url{https://link.aps.org/doi/10.1103/PhysRevLett.117.087402}

\bibitem{McIver2020}
McIver J~W, Schulte B, Stein F~U, Matsuyama T, Jotzu G, Meier G and Cavalleri A
  2020 {\em Nat. Phys.\/} {\bf 16} 38--41 ISSN 1745-2481
  \urlprefix\url{https://doi.org/10.1038/s41567-019-0698-y}

\bibitem{Fan2024}
Fan B, Duan W, Rubio A and Tang P 2024 {\em npj Quantum Materials\/} {\bf 9}
  101 ISSN 2397-4648 \urlprefix\url{https://doi.org/10.1038/s41535-024-00714-7}

\bibitem{PhysRevLett.126.196402}
Zhao Y~X, Chen C, Sheng X~L and Yang S~A 2021 {\em Phys. Rev. Lett.\/} {\bf
  126}(19) 196402
  \urlprefix\url{https://link.aps.org/doi/10.1103/PhysRevLett.126.196402}

\bibitem{PhysRevLett.130.026101}
Meng Y, Lin S, Shi B~j, Wei B, Yang L, Yan B, Zhu Z, Xi X, Wang Y, Ge Y, Yuan
  S~q, Chen J, Liu G~G, Sun H~x, Chen H, Yang Y and Gao Z 2023 {\em Phys. Rev.
  Lett.\/} {\bf 130}(2) 026101
  \urlprefix\url{https://link.aps.org/doi/10.1103/PhysRevLett.130.026101}

\bibitem{PhysRevLett.123.256402}
Sheng X~L, Chen C, Liu H, Chen Z, Yu Z~M, Zhao Y~X and Yang S~A 2019 {\em Phys.
  Rev. Lett.\/} {\bf 123}(25) 256402
  \urlprefix\url{https://link.aps.org/doi/10.1103/PhysRevLett.123.256402}

\bibitem{PhysRevB.90.125143}
Asb\'oth J~K, Tarasinski B and Delplace P 2014 {\em Phys. Rev. B\/} {\bf
  90}(12) 125143
  \urlprefix\url{https://link.aps.org/doi/10.1103/PhysRevB.90.125143}

\bibitem{PhysRevB.96.195303}
Yao S, Yan Z and Wang Z 2017 {\em Phys. Rev. B\/} {\bf 96}(19) 195303
  \urlprefix\url{https://link.aps.org/doi/10.1103/PhysRevB.96.195303}

\bibitem{PhysRevB.103.115308}
Nag T, Juri\ifmmode \check{c}\else \v{c}\fi{}i\ifmmode~\acute{c}\else
  \'{c}\fi{} V and Roy B 2021 {\em Phys. Rev. B\/} {\bf 103}(11) 115308
  \urlprefix\url{https://link.aps.org/doi/10.1103/PhysRevB.103.115308}

\bibitem{PhysRevX.3.031005}
Rudner M~S, Lindner N~H, Berg E and Levin M 2013 {\em Phys. Rev. X\/} {\bf
  3}(3) 031005
  \urlprefix\url{https://link.aps.org/doi/10.1103/PhysRevX.3.031005}

\bibitem{RevModPhys.89.011004}
Eckardt A 2017 {\em Rev. Mod. Phys.\/} {\bf 89}(1) 011004
  \urlprefix\url{https://link.aps.org/doi/10.1103/RevModPhys.89.011004}

\bibitem{PhysRevLett.128.127601}
Benalcazar W~A and Cerjan A 2022 {\em Phys. Rev. Lett.\/} {\bf 128}(12) 127601
  \urlprefix\url{https://link.aps.org/doi/10.1103/PhysRevLett.128.127601}

\bibitem{PhysRevLett.118.056401}
Zhao Y~X and Lu Y 2017 {\em Phys. Rev. Lett.\/} {\bf 118}(5) 056401
  \urlprefix\url{https://link.aps.org/doi/10.1103/PhysRevLett.118.056401}

\bibitem{PhysRevB.103.045424}
Ghosh A~K, Nag T and Saha A 2021 {\em Phys. Rev. B\/} {\bf 103}(4) 045424
  \urlprefix\url{https://link.aps.org/doi/10.1103/PhysRevB.103.045424}

\bibitem{PhysRevLett.121.076802}
Yates D, Lemonik Y and Mitra A 2018 {\em Phys. Rev. Lett.\/} {\bf 121}(7)
  076802
  \urlprefix\url{https://link.aps.org/doi/10.1103/PhysRevLett.121.076802}

\bibitem{PhysRevB.105.L081102}
Du X~L, Chen R, Wang R and Xu D~H 2022 {\em Phys. Rev. B\/} {\bf 105}(8)
  L081102 \urlprefix\url{https://link.aps.org/doi/10.1103/PhysRevB.105.L081102}

\bibitem{PhysRevB.103.L041115}
Wu H, Wang B~Q and An J~H 2021 {\em Phys. Rev. B\/} {\bf 103}(4) L041115
  \urlprefix\url{https://link.aps.org/doi/10.1103/PhysRevB.103.L041115}

\bibitem{PhysRevB.104.205117}
Wang B~Q, Wu H and An J~H 2021 {\em Phys. Rev. B\/} {\bf 104}(20) 205117
  \urlprefix\url{https://link.aps.org/doi/10.1103/PhysRevB.104.205117}

\bibitem{PhysRevA.100.023622}
Liu H, Xiong T~S, Zhang W and An J~H 2019 {\em Phys. Rev. A\/} {\bf 100}(2)
  023622 \urlprefix\url{https://link.aps.org/doi/10.1103/PhysRevA.100.023622}

\bibitem{PhysRevB.102.041119}
Wu H and An J~H 2020 {\em Phys. Rev. B\/} {\bf 102}(4) 041119(R)
  \urlprefix\url{https://link.aps.org/doi/10.1103/PhysRevB.102.041119}

\bibitem{PhysRevB.107.035128}
Gao M~J, Wu H and An J~H 2023 {\em Phys. Rev. B\/} {\bf 107}(3) 035128
  \urlprefix\url{https://link.aps.org/doi/10.1103/PhysRevB.107.035128}

\bibitem{PhysRevB.109.184518}
Gao M~J, Ma Y~P and An J~H 2024 {\em Phys. Rev. B\/} {\bf 109}(18) 184518
  \urlprefix\url{https://link.aps.org/doi/10.1103/PhysRevB.109.184518}

\bibitem{PhysRevB.110.125427}
Ghosh A~K, Nag T and Saha A 2024 {\em Phys. Rev. B\/} {\bf 110}(12) 125427
  \urlprefix\url{https://link.aps.org/doi/10.1103/PhysRevB.110.125427}

\bibitem{PhysRevB.107.235132}
Wu H and An J~H 2023 {\em Phys. Rev. B\/} {\bf 107}(23) 235132
  \urlprefix\url{https://link.aps.org/doi/10.1103/PhysRevB.107.235132}

\bibitem{PhysRevB.102.125144}
Xiong Z, Lin Z~K, Wang H~X, Zhang X, Lu M~H, Chen Y~F and Jiang J~H 2020 {\em
  Phys. Rev. B\/} {\bf 102}(12) 125144
  \urlprefix\url{https://link.aps.org/doi/10.1103/PhysRevB.102.125144}

\bibitem{xz4g-jlbg}
Wu H and An J~H 2025 {\em Phys. Rev. B\/} {\bf 112}(4) L041101
  \urlprefix\url{https://link.aps.org/doi/10.1103/xz4g-jlbg}

\bibitem{PhysRevLett.132.197202}
Xiang X, Peng Y~G, Gao F, Wu X, Wu P, Chen Z, Ni X and Zhu X~F 2024 {\em Phys.
  Rev. Lett.\/} {\bf 132}(19) 197202
  \urlprefix\url{https://link.aps.org/doi/10.1103/PhysRevLett.132.197202}

\bibitem{PhysRevLett.133.176602}
Han Y, Cui C, Li X~P, Zhang T~T, Zhang Z, Yu Z~M and Yao Y 2024 {\em Phys. Rev.
  Lett.\/} {\bf 133}(17) 176602
  \urlprefix\url{https://link.aps.org/doi/10.1103/PhysRevLett.133.176602}

\bibitem{Pan2023}
Pan Y, Cui C, Chen Q, Chen F, Zhang L, Ren Y, Han N, Li W, Li X, Yu Z~M, Chen H
  and Yang Y 2023 {\em Nature Communications\/} {\bf 14} 6636 ISSN 2041-1723
  \urlprefix\url{https://doi.org/10.1038/s41467-023-42457-2}

\bibitem{PhysRevLett.131.157201}
Wang D, Deng Y, Ji J, Oudich M, Benalcazar W~A, Ma G and Jing Y 2023 {\em Phys.
  Rev. Lett.\/} {\bf 131}(15) 157201
  \urlprefix\url{https://link.aps.org/doi/10.1103/PhysRevLett.131.157201}

\bibitem{PhysRevB.108.205135}
Li Y, Qiu H, Zhang Q and Qiu C 2023 {\em Phys. Rev. B\/} {\bf 108}(20) 205135
  \urlprefix\url{https://link.aps.org/doi/10.1103/PhysRevB.108.205135}

\bibitem{PhysRevApplied.20.064042}
Li Y, Zhang J~H, Mei F, Xie B, Lu M~H, Ma J, Xiao L and Jia S 2023 {\em Phys.
  Rev. Appl.\/} {\bf 20}(6) 064042
  \urlprefix\url{https://link.aps.org/doi/10.1103/PhysRevApplied.20.064042}

\bibitem{Wintersperger2020}
Wintersperger K, Braun C, Ünal F~N, Eckardt A, Liberto M~D, Goldman N, Bloch I
  and Aidelsburger M 2020 {\em Nat. Phys.\/} {\bf 16} 1058 ISSN 1745-2481
  \urlprefix\url{https://doi.org/10.1038/s41567-020-0949-y}

\bibitem{Maczewsky2017}
Maczewsky L~J, Zeuner J~M, Nolte S and Szameit A 2017 {\em Nat. Commun.\/} {\bf
  8} 13756 ISSN 2041-1723 \urlprefix\url{https://doi.org/10.1038/ncomms13756}

\bibitem{PhysRevLett.122.173901}
Cheng Q, Pan Y, Wang H, Zhang C, Yu D, Gover A, Zhang H, Li T, Zhou L and Zhu S
  2019 {\em Phys. Rev. Lett.\/} {\bf 122}(17) 173901
  \urlprefix\url{https://link.aps.org/doi/10.1103/PhysRevLett.122.173901}

\bibitem{PhysRevLett.129.254301}
Cheng Z, Bomantara R~W, Xue H, Zhu W, Gong J and Zhang B 2022 {\em Phys. Rev.
  Lett.\/} {\bf 129}(25) 254301
  \urlprefix\url{https://link.aps.org/doi/10.1103/PhysRevLett.129.254301}

\bibitem{PhysRevLett.133.073803}
Lin Z, Song W, Wang L~W, Xin H, Sun J, Wu S, Huang C, Zhu S, Jiang J~H and Li T
  2024 {\em Phys. Rev. Lett.\/} {\bf 133}(7) 073803
  \urlprefix\url{https://link.aps.org/doi/10.1103/PhysRevLett.133.073803}

\end{thebibliography}

\end{document}